\documentclass[12pt]{article}
 \usepackage{inchmargs}
 \usepackage{splat}
 \usepackage{hyperref}

\makeatletter

\newcommand{\bibintro}[1]{%
\let\@origthebib=\thebibliography
\let\@origlist=\list
\def\thebibliography{\def\list{#1\par\bigskip\@origlist}\@origthebib}%
}

\makeatother

\newcommand{\inlineciteclosed}
  {\ifx\newblock\undefined\newcommand{\newblock}{\hskip .11em plus .33em minus .07em}\fi}

\inlineciteclosed

\newcounter{oscounter}

\newcommand{\ordinal}[1]{#1\ordsuffix{#1}}

\newcommand{\ordsuffix}[1] 
{\ifcase#1
th\or st\or nd\or rd\or th\or th\or th\or th\or th\or th\or
th\or th\or th\or th\else\setcounter{oscounter}{#1}\ordsuffixrecur\fi}

\newcommand{\ordsuffixrecur}
{\addtocounter{oscounter}{-10}\ifcase\value{oscounter}%
th\or st\or nd\or rd\or th\or th\or th\or th\or th\or th\else
\ordsuffixrecur\fi}

\newcommand{\soda}[1]
    {Proceedings of the \ordinal{#1} Annual {\ifnum#1>3ACM-\fi}SIAM
Symposium on Discrete Algorithms}

\newcommand{\spaa}[1]
    {Proceedings of the
     \ifcase#1\or 1989 \else{\ordinal{#1} Annual }\fi
     ACM Symposium on Parallel Algorithms and Architectures}

\newcommand{\rigaddr}{Dept.\ of Mathematical and Computer Sciences\\
 Loyola University\\
 6525 N. Sheridan Rd.\\
 Chicago, IL\ \ 60626-5385}

\newcommand{\binomial}[2]{{{#1}\atopwithdelims(){#2}}}
\newcommand{\floor}[1]{\left\lfloor#1\right\rfloor}
\newcommand{\ceil}[1]{\left\lceil#1\right\rceil}

\newcommand{\A}{{\tt a}}
\newcommand{\B}{{\tt b}}
\newcommand{\C}{{\tt c}}
\newcommand{\D}{{\tt d}}
\newcommand{\X}{{\tt x}}

\newcommand{\half}{\frac{1}{2}}
\newcommand{\goldvar}{\half\left(1+1/\sqrt{5}\,\right)}
\newcommand{\goldvarconj}{\half\left(1-1/\sqrt{5}\,\right)}

 \newenvironment{proof}{\noindent {\it Proof.}\ \ }{}
\makeatletter
\def\qed{
   {%
      \unskip
      \nobreak \hfil
      \penalty 50               
      \hskip 3em                
      \null \nobreak \hfil
      \qedbox
      \parfillskip=\z@skip
      \finalhyphendemerits=\z@  
      \endgraf                  
   }}
\makeatother
\newcommand{\qedbox}{\ \hfill\rule{1.5ex}{1.5ex}\vspace{\smallskipamount}}

\newtheorem{theorem}{Theorem}
\newtheorem{corollary}[theorem]{Corollary}
\newtheorem{lemma}[theorem]{Lemma}

\begin{document}

\date{} 

\title{Bounds on the Number of Longest Common Subsequences}

\author{Ronald I. Greenberg\\
 \rigaddr\\
 \href{mailto:rig@cs.luc.edu}{{\tt rig@cs.luc.edu}}, \url{http://www.cs.luc.edu/~rig}}

\maketitle

\begin{abstract}
 This paper performs the analysis necessary to bound the running time
of known, efficient algorithms for generating all longest common
subsequences.  That is, we bound the running time as a function of
input size for algorithms with time essentially proportional to the
output size.  This paper considers both the case of computing all
distinct LCSs and the case of computing all LCS embeddings.  Also
included is an analysis of how much better the efficient algorithms
are than the standard method of generating LCS embeddings.  A full
analysis is carried out with running times measured as a function of
the total number of input characters, and much of the analysis is also
provided for cases in which the two input sequences are of the same
specified length or of two independently specified lengths.

\end{abstract}


\section{Background and Terminologies}
 \label{sec:intro}
 Let $A=a_1a_2\ldots a_m$ and $B=b_1b_2\ldots b_n$ ($m\leq n$) be
two sequences over an alphabet $\Sigma$.  A sequence that can be
obtained by deleting some symbols of another sequence is referred to
as a {\em subsequence} of the original sequence.  A {\em common
subsequence} of $A$ and $B$ is a subsequence of both $A$ and $B$.  A
longest common subsequence (LCS) is a common subsequence of greatest
possible length.  A pair of sequences may have many different LCSs.
In addition, a single LCS may have many different {\em embeddings},
i.e., positions in the two strings to which the characters of the LCS
correspond.

Most investigations of the LCS problem have focused on efficiently
finding one LCS.  A widely familiar $O(mn)$ dynamic programming
approach goes back at least as far as the early
1970s~\cite{NeedlemanW1970,Sankoff1972,WagnerF1974}, and many later
studies have focused on improving the time and/or space required for
the computation
(e.g.~\cite{HuntS1977,Hirschberg1977,MasekP1980,NakatsuKY1982,HsuD1984,Ukkonen1985,Apostolico1986,Myers1986,ApostolicoG1987,KumarR1987,WuMMM1990,ChinP1990,ApostolicoBG1992,EppsteinGGI1992,Rick2000S,GoemanC2002}.).

The familiar dynamic programming approach provides a basis for
generating {\em all}\/ LCSs, but the naive approach
(e.g.~\cite{AhoU1995}) may generate the same LCS or even the same LCS
embedding many times.  Other methods have been developed to generate a
listing of all distinct LCSs or all LCS embeddings in time
proportional to the output size (plus a preprocessing time of $O(mn)$
or less), i.e., without generating
duplicates~\cite{AltschulE1986,Gotoh1990,Rick2000E,Greenberg2002F}.
But prior works give no indication of how long the running time may be
as a function of input size.  They also do not indicate how the
asymptotic time of the efficient methods compares to the naive
approach, so it is unclear how worthwhile it is to implement the more
complex algorithms.

Section~\ref{sec:numlcs} of this paper obtains bounds on the amount of
time that may be required to find all distinct LCSs of two input
sequences of fixed total length when using an algorithm with time
proportional to the output size.  Technically, the time is governed by
the LCS length times the number of LCSs, but we focus on bounding just
the maximum possible number of distinct LCSs.  (There will be little
difference in the LCS lengths that maximize these two measures.)

Section~\ref{sec:embeddings} similarly bounds the amount of time that
may be required to find all LCS embeddings.  Here an exact computation
of the maximum possible number of LCS embeddings is provided, and the
analysis is carried out both for a fixed total number of input
characters and for two input sequences of the same fixed length.  In
addition, a partial analysis is provided for two input sequences with
independently specified lengths.  Since the maximum number of LCS
embeddings is achievable when there is just one distinct LCS, the
results in this section also give a measure of how much more efficient
it is to generate all distinct LCSs in time proportional to the output
size, as compared to a method that efficiently generates all
embeddings and removes duplicate LCSs.

Section~\ref{sec:naive} indicates how much more efficient a fast
algorithm for generating all LCS embeddings (or all distinct LCSs) may
be in comparison to the standard method that may even report the same
embedding more than once.  It turns out that the naive algorithm may
even generate the same embedding of a single LCS exponentially many
times, and we precisely quantify the asymptotic worst-case overhead.

\section{Bounding the Number of Distinct LCSs}
 \label{sec:numlcs}
 In this section we determine how much time an efficient algorithm for
listing all distinct LCSs may require.  While the actual time is the
LCS length $l$ times the number of distinct LCSs (plus preprocessing
time), we focus here on bounding just the maximum number of distinct
LCSs.  The values of $l$ that maximize these two measures will be
increasingly close as the sizes of the input sequences (and
consequently $l$) grow.  Throughout this section, we will let $D(t)$
denote the maximum possible number of distinct LCSs for two input
sequences of total length $t$ (assuming an unbounded alphabet).

Letting $m=\floor{t/2}$ and $z=(-\floor{t/2})\bmod3$, a lower bound on
$D(t)$ follows from considering two input sequences of length $m$ of
the form
 $X\!\mathtt{efghijklm...}$
and
 $Y\!\mathtt{gfejihmlk...}$,
 where $X$ and $Y$ are empty if $z=0$, {\tt ab} and {\tt ba} if $z=1$,
or {\tt abcd} and {\tt badc} if $z=2$:


\begin{theorem}
 \label{thm:distinct-lower}
 For $t\geq4$,
 $D(t)\geq3^{(\floor{t/2}-2((-\floor{t/2})\bmod3))/3}2^{(-\floor{t/2})\bmod3}$ (which implies that
 for $t$ divisible by 6, $D(t)\geq3^{t/6}>1.2^t$).
 \qed
 \end{theorem}

To obtain an upper bound, we begin with the following lemma.  For this
purpose, we define an embedding of a character of an LCS in the two
input sequences $A$ and $B$ as an ordered pair of a position in $A$
and a position in $B$ from which the character may be selected when
forming the LCS.  We say that two character embeddings $(p,q)$ and
$(p',q')$ {\em cross} if $p<p'\wedge q'<q$ or $p'<p\wedge q<q'$.

\begin{lemma}
 \label{lem:distinct-upper}
 Consider two LCSs starting with different characters and any embedding
of these two LCSs.  The embeddings of the initial characters of these
two LCSs must cross.
 \end{lemma}

\begin{proof}
 Suppose there are embeddings of two LCSs $C=c_1c_2\ldots c_l$ and
$C'=c'_1c'_2\ldots c'_l$ such that $c_1\not=c'_1$ and the embeddings
of $c_1$ and $c'_1$ do not cross.  Then $cC'$ or $c'C$ is a common
subsequence, contradicting the assumption that $C$ and $C'$ are LCSs.
 \qed
 \end{proof}

\begin{theorem}
 \label{thm:distinct-upper}
 $D(t)\leq4^{t/5}<1.32^t$.
 \end{theorem}

\begin{proof}
 The proof is by induction on $t$; the base case is easy to check.
For the induction step, let $k$ be the number of choices for the first
character when constructing an LCS from the two given input strings.
Since the embeddings of all these possible initial characters must
cross, the sum of the two string positions corresponding to each such
embedding must be at least $k+1$.  Furthermore, once such an embedding
is chosen for the first character of the LCS, $k+1$ characters of the
input strings are removed from consideration for construction of the
rest of the LCS, since no other character of the LCS can have an
embedding that crosses the first one.  Thus, $D(t)\leq kD(t-(k+1))$.
By the induction hypothesis, $D(t)\leq k4^{(t-(k+1))/5}$.  Then the
result follows from the observation that $k/4^{(k+1)/5}$ is decreasing
for $k>5/\ln4\approx3.6$ and is at most 1 for integral $k\leq4$.
 \qed
 \end{proof}

Neither Theorem~\ref{thm:distinct-lower} nor
Theorem~\ref{thm:distinct-upper} is tight; e.g., with $t=10$, there
are 7 distinct LCSs for the input strings $\A\B\C\D\A$ and
$\C\B\A\D\C$.  If neither input string contains repeated characters,
however, we can combine Theorem~\ref{thm:distinct-lower} with the
following theorem to obtain tight upper and lower bounds:

\begin{theorem}
 \label{thm:distinct-upper-norepeats}
 $D(t)\leq3^{(\floor{t/2}-2((-\floor{t/2})\bmod3))/3}2^{(-\floor{t/2})\bmod3}$
if there are no repeated characters in either input sequence.
 \qed
 \end{theorem}

\begin{proof}
 We proceed by induction as in Theorem~\ref{thm:distinct-upper}, but
now when we make one of $k$ choices for the first character of the
LCS, we eliminate $2k$ characters from possible use in the rest of the
LCS.  Thus, $D(t)\leq kD(t-2k)$.  Using the induction hypothesis and
considering the different cases for the values of $\floor{t/2}$ and
$k$ modulo 3, the result follows as long as we can show that
 $k3^{(-k+2((-k)\bmod3))/3}2^{-((-k)\bmod3)}\leq1$.  We succeed
by checking $k=1$, 2, 3, 4, and 5 and noting that $k/3^{k/3}$ is
decreasing for $k\geq3/\ln3\approx2.7$.
 \qed
 \end{proof}

Note that the results in this section also apply in the case that we
require $m=n$, by setting $t=2n$.

\section{The Maximum number of LCS Embeddings}
 \label{sec:embeddings}
 In this section, we determine how much time an efficient algorithm for
listing all LCS embeddings may require.  Utilizing the same
justification as in the previous section, we neglect LCS length as a
component of the running time.  (It would actually be easy to
incorporate LCS length into the presentation in this section, and this
may be done in the full paper.)  Thus we focus on computing the
maximum possible number of LCS embeddings.  In the full paper, we will
argue that the maximum number of LCS embeddings can be achieved when
there is just one distinct LCS.  Therefore, we turn our attention to
computing the maximum possible number of embeddings of a single LCS.
This result will also indicate how much more efficient it is to
generate all distinct LCS embeddings in time proportional to output
size rather than to generate all embeddings (efficiently) and remove
duplicates.

We begin by determining the maximum number of embeddings of an LCS of
length $l$ in two input strings of length $m$ and $n$.  Then we
perform the maximization over $l$ in the cases of (1) $m=n$ and (2)
$m$ and $n$ variable but with $m+n$ fixed at $t$.

\begin{lemma}
 \label{lem:embeddings-l-maximization}
 The maximum possible number of embeddings $E(n,m,l)$ of a single LCS
of length $l$ in two input sequences of lengths $m$ and $n$ is
 $$ E(n,m,l) = \max_{y\leq l} \binomial{m-y}{l-y}\binomial{n+y-l}{y} \ . $$
 \end{lemma}

\begin{proof}
 First,
 $E(n,m,l) \geq \max_{y\leq l} \binomial{m-y}{l-y}\binomial{n+y-l}{y}$,
 because, for any $y\leq l$, we can find
 $\binomial{m-y}{l-y}\binomial{n+y-l}{y}$ embeddings of the string
 $\A^{l-y}\B^y$ in the two strings 
 $\A^{m-y}\B^y$ and
 $\A^{l-y}\B^{n+y-l}$
 (where $\X^n$ represents $n$ repetitions of the
character $\X$).

 Now we prove
 $E(n,m,l) \leq \max_{y\leq l} \binomial{m-y}{l-y}\binomial{n+y-l}{y}$
as follows.  Each character of any LCS must have a fixed embedding in
at least one of the two input strings
 $A=a_1a_2\ldots a_m$ and
 $B=b_1b_2\ldots b_n$.  (Suppose, to the contrary, that $c_k$ of the LCS
 $C=c_1c_2\ldots c_l$ could be embedded into $a_i$ or $a_j$ ($i<j$) and into
 $b_p$ or $b_q$ ($p<q$).  Then
 $c_1c_2\ldots c_{k-1}$ could be embedded in
 $a_1a_2\ldots a_{i-1}$ and in
 $b_1b_2\ldots b_{p-1}$, while
 $c_{k+1}c_{k+2}\ldots c_l$ could be embedded in
 $a_{j+1}a_{j+2}\ldots a_m$ and in
 $b_{q+1}b_{q+2}\ldots b_n$.  This contradicts the supposition that
$C$ is an LCS, because we now know that
  $c_1c_2\ldots c_{k-1}c_kc_kc_{k+1}c_{k+2}\ldots c_l$ is a common
subsequence of $A$ and $B$.)  Let $y$ be the number of characters of
the LCS under consideration that have a fixed embedding in $A$.  Then
at least $l-y$ characters have a fixed embedding in $B$.  Now the
number of ways to embed those $l-y$ characters in $A$ is at most
$\binomial{m-y}{l-y}$, and the number of ways to embed into $B$ the
$y$ characters fixed in $A$ is at most $\binomial{n-(l-y)}{y}$.
 \qed
 \end{proof}

\begin{lemma}
 \label{lem:embeddings-l}
 The maximum possible number of embeddings $E(n,m,l)$ of a single LCS
of length $l$ in two input sequences of lengths $m$ and $n$ is
 $$ E(n,m,l) = \binomial{m-y^*}{l-y^*}\binomial{n+y^*-l}{y^*} \ , $$
 where
 $$ y^* = \ceil{\frac{l(n-l)+l-m}{m+n-2l}} \ . $$
 \end{lemma}

\begin{proof}
 The result follows from Lemma~\ref{lem:embeddings-l-maximization} as
long as we can show that $y^*$ (which satisfies $y^*\leq l$) is the
(nonnegative integral) value of $y$ that maximizes
 $P(y)=\binomial{m-y}{l-y}\binomial{n+y-l}{y}$.  To do this, we show
that $P(y+1)\leq P(y)$ if and only if
 $y\geq\frac{l(n-l)+l-m}{m+n-2l}$ as follows:
 \begin{eqnarray*}
 P(y+1)\leq P(y) &\iff& \frac{(m-y-1)!}{(l-y-1)!\,(m-l)!}\frac{(n+y+1-l)!}{(y+1)!\,(n-l)!}
                        \leq\frac{(m-y)!}{(l-y)!\,(m-l)!}\frac{(n+y-l)!}{y!\,(n-l)!} \\
    &\iff& (l-y)(n+y-l+1)\leq(m-y)(y+1) \\
    &\iff& -y^2+y(l-n+l-1)+ln-l^2+l \leq -y^2+y(m-1)+m \\
    &\iff& l(n-l)+l-m \leq y(m+n-2l)
 \end{eqnarray*}
 (Note that $m+n-2l\geq0$, and where it equals 0 ($n=m=l$), the result
remains correct as long as we use the convention of interpreting
$\frac{0}{0}$ as 1.)
 \qed
 \end{proof}

Next we specialize to $m=n$.

\begin{lemma}
 \label{lem:embeddings-l-m=n}
 The maximum possible number of embeddings $E(n,n,l)$ of a single LCS
of length $l$ in two input sequences of length $n$ is
 $$ E(n,n,l) = \binomial{n-\floor{l/2}}{\ceil{l/2}}\binomial{n-\ceil{l/2}}{\floor{l/2}}
    = \binomial{n-\floor{l/2}}{n-l}\binomial{n-\ceil{l/2}}{n-l} \ . $$
 \end{lemma}

\begin{proof}
 Substituting $n$ for $m$ in Lemma~\ref{lem:embeddings-l} gives
$y^*=\ceil{(l-1)/2}=\floor{l/2}$.  Then substituting for $m$ and $y^*$
in the expression for $E(n,m,l)$ there gives the desired result after
using basic facts about floors and ceilings and the relationship
$\binomial{r}{k}=\frac{r!}{k!\,(r-k)!}$.
 \qed
 \end{proof}

\begin{lemma}
 \label{lem:embeddings-m=n}
 Let $\sigma=(5n-1-\sqrt{5(n+1)^2-4})/5$ and $\tau=(5n-\sqrt{5(n+1)^2})/5$.
 Then, the maximum possible number of embeddings of a single LCS in
two input sequences of length $n$
 is achieved with an LCS length $l^*$ that satisfies the following conditions.
 \begin{enumerate}
 \item $l^*$ can be chosen as either $\sigma$ or $\sigma+1$ if $\sigma$ is integral.
 \item Otherwise, $l^*=\ceil{\sigma}$ if $\ceil{\sigma}$ is even.
 \item Otherwise, $l^*=\ceil{\tau}$ (which most often equals $\ceil{\sigma}$).
 \end{enumerate}
 \end{lemma}

\begin{proof}
 From Lemma~\ref{lem:embeddings-l-m=n}, we see
 that the condition $E(n,n,l+1)\leq E(n,n,l)$ is equivalent to:
 \begin{eqnarray*}
 \frac{\left(n-\floor{\frac{l+1}{2}}\right)!}{\ceil{\frac{l+1}{2}}!\,(n-l-1)!}
 \frac{\left(n-\ceil{\frac{l+1}{2}}\right)!}{\floor{\frac{l+1}{2}}!\,(n-l-1)!}
    &\leq& 
    \frac{\left(n-\floor{\frac{l}{2}}\right)!}{\ceil{\frac{l}{2}}!\,(n-l)!}
    \frac{\left(n-\ceil{\frac{l}{2}}\right)!}{\floor{\frac{l}{2}}!\,(n-l)!}
    \ i.e., \\
 (n-l)^2 &\leq& \ceil{\frac{l+1}{2}}(n-\floor{l/2}) \\
    && \mbox{since $\floor{\frac{l+1}{2}}=\ceil{\frac{l}{2}}$,
                   and $\ceil{\frac{l+1}{2}}=\floor{\frac{l}{2}}+1$} \ .
 \end{eqnarray*}
 Now we see that the condition $E(n,n,l+1)\leq E(n,n,l)$ is equivalent to:
 \begin{eqnarray}
 \label{ineq:optleven}
 5l^2-2(5n-1)l+4n(n-1) &\leq& 0 \qquad\mbox{for $l$ even} \\
 \label{ineq:optlodd}
 5l^2-10nl+4n^2-2n-1 &\leq& 0 \qquad\mbox{for $l$ odd} \ .
 \end{eqnarray}
 For each of these two quadratic expressions in $l$, the roots $r_1$
and $r_2$ satisfy $-1<r_1<n$ and $n<r_2$.  Since $l\leq n$, in each of
the cases of $l$ even and $l$ odd, an (integral) value of $l$
maximizing $E(n,n,l)$ is $\ceil{r_1}$ as long as that value is of the
appropriate parity.  The values of $r_1$ are $\sigma$ and $\tau$, in~(\ref{ineq:optleven}) and~(\ref{ineq:optlodd}), respectively, and it may
be noted that $\tau$ is never integral, since $\sqrt{5}$ is
irrational.

Now, if $\sigma$ is an even integer, we see that according
to~(\ref{ineq:optleven}), $l^*$ is selectable as $\sigma$ or $\sigma+1$.
Furthermore, this is the final result, since the odd value
$\ceil{\tau}$ that maximizes $E(n,n,l)$ according
to~(\ref{ineq:optlodd}) is equal to $\sigma+1$.

It is easy to check that any time $\sigma$ is integral, it is even, so
it only remains to consider $\sigma$ nonintegral.  We see that if
$\ceil{\sigma}$ is even, then according to~(\ref{ineq:optleven}),
$l^*=\sigma$.  Furthermore,~(\ref{ineq:optlodd}) will not lead to a
better value of $E(n,n,l)$, since $\sigma<\tau<\sigma+\frac{1}{5}$.
Finally, if $\ceil{\sigma}$ is odd, then either $\ceil{\tau}$ has the
same odd value, or $\ceil{\tau}$ has the even value $\ceil{\sigma}+1$;
either way, we see from~(\ref{ineq:optleven}) and~(\ref{ineq:optlodd})
that $l^*=\ceil{\tau}$.
 \qed
 \end{proof}

\begin{theorem}
 \label{thm:embeddings-m=n}
 The maximum possible number of embeddings of a single LCS in two
input sequences of length $n$ is
 $$ \binomial{\floor{\goldvar(n+1)}}{\ceil{\left(5n-1-\sqrt{5(n+1)^2-4}\,\right)/10}}
    \binomial{\floor{\left(5n+1+\sqrt{5(n+1)^2-4}\,\right)/10}}{\floor{\goldvarconj(n+1)}} \ . $$
\end{theorem}

 \begin{proof}
 The result follows from Lemma~\ref{lem:embeddings-m=n} as follows.
 Since $\sigma<\tau<\sigma+1$, we know $\floor{\ceil{\tau}/2}=\floor{\ceil{\sigma}/2}$ if
$\ceil{\sigma}$ is even.  Thus, by Lemma~\ref{lem:embeddings-m=n},
 $\floor{l^*/2}=\floor{\ceil{\tau}{2}}$.
 Similarly, there is an acceptable $l^*$ with
 $\ceil{l^*/2}=\ceil{\ceil{\sigma}/2}$.  Now we have the following four relationships:
 \begin{itemize}
 \item $\ceil{l^*/2}=\ceil{\ceil{\sigma}/2}=\ceil{\sigma/2}$
 \item $n-\ceil{l^*/2}=n-\ceil{\sigma/2}=\floor{n-\sigma/2}$
 \item $\floor{l^*/2}=\floor{\ceil{\tau}/2}=\floor{\floor{(\tau+1)/2}}=\floor{(\tau+1)/2}$
 \item $n-\floor{l^*/2}=n-\floor{(\tau+1)/2}=n-\ceil{(\tau-1)/2}=\floor{n-(\tau-1)/2}$
 \end{itemize}
 Substituting these four relationships into
 $E(n,n,l)=\binomial{n-\floor{l/2}}{\ceil{l/2}}\binomial{n-\ceil{l/2}}{\floor{l/2}}$
 from Lemma~\ref{lem:embeddings-l-m=n} yields the desired result.
 \qed
 \end{proof}

\begin{corollary}
 \label{cor:embeddings-m=n}
 The limit as $n$ goes to infinity of the maximum possible number of
embeddings of a single LCS in two input sequences of length $n$ is
 $$ \frac{\phi^2\sqrt{5}}{2\pi}\left(\phi^2\right)^n\!\bigg/n \approx .932(2.62)^n\!/n \ , $$
 where $\phi=(1+\sqrt{5})/2$ (the golden ratio).
 \end{corollary}

\begin{proof}
 We will use Stirling's approximation to the factorial:
 \begin{eqnarray}
 \label{eqn:Stirling}
 n!=\sqrt{2\pi n}(n/e)^n(1+\Theta(1/n))\qquad\mbox{\cite[p.\ 111]{Knuth1973}} \ .
 \end{eqnarray}
 Then, the limit as $n$ goes to infinity of the expression in
Theorem~\ref{thm:embeddings-m=n} is:
 \begin{eqnarray*}
 \lim_{n\rightarrow\infty}\binomial{\goldvar n}{\goldvarconj n}^2
 &=&\lim_{n\rightarrow\infty}\binomial{\phi n/\sqrt{5}}{(\phi-1)n/\sqrt{5}}^2\\
 &=&\lim_{n\rightarrow\infty}\left(\frac{\left(\phi n/\sqrt{5}\,\right)!}{\left((\phi-1)n/\sqrt{5}\,\right)!\,\left(n/\sqrt{5}\,\right)!}\right)^2 \\
 &=& \left(
     \sqrt{\frac{\phi\sqrt{5}}{(\phi-1)2\pi n}}\,
     \left(\phi^{\phi n/\sqrt{5}}\!\bigg/(\phi-1)^{(\phi-1)n/\sqrt{5}}\,\right)
     \right)^2 \qquad\mbox{by Eqn.~\ref{eqn:Stirling}}\\
 &=& \left(
     \sqrt{\frac{\phi^2\sqrt{5}}{2\pi n}}\,
     \left(\phi^{\phi n/\sqrt{5}}\phi^{(\phi-1)n/\sqrt{5}}\,\right)
     \right)^2
 \end{eqnarray*}
 \qed
 \end{proof}

Next, we consider the case in which the total number of characters in
the two input strings is fixed, but the lengths of the individual
strings are not.  The following Lemma follows immediately from
Lemma~\ref{lem:embeddings-l-maximization}, using the fact
 $\binomial{r}{k}\binomial{r'}{k'}\leq\binomial{r+r'}{k+k'}$

\begin{lemma}
 \label{lem:embeddings-l-t}
 The maximum possible number of embeddings of a single LCS
of length $l$ in two input sequences of total length $t$ is
 $ \binomial{t-l}{l} \ . $
 \qed
 \end{lemma}

Using Lemma~\ref{lem:embeddings-l-t} and reasoning similar to the
proof of Lemma~\ref{lem:embeddings-m=n}, yields the following theorem.

\begin{theorem}
 \label{thm:embeddings-t}
 The maximum possible number of embeddings of a single LCS in two
input sequences of total length $t$ is
 $$ \binomial{\floor{\left(5t+3+\sqrt{5(t+1)^2+4}\,\right)/10}}
              {\ceil{\left(5t-3-\sqrt{5(t+1)^2+4}\,\right)/10}} \ . $$
 \end{theorem}


Finally, we can use the same type of reasoning as in
Corollary~\ref{cor:embeddings-m=n} to reach the following conclusion:

\begin{corollary}
 \label{cor:embeddings-t}
 The limit as $t$ goes to infinity of the maximum possible number of
embeddings of a single LCS in two input sequences of total length $t$
is
 $ \phi\sqrt{\sqrt{5}/(2\pi)}\,\phi^t\!/\sqrt{t} \approx .965(1.62)^t\!/\sqrt{t} \ . $
 \qed
 \end{corollary}

\section{The Degree of Inefficiency of Naively Generating all LCSs}
 \label{sec:naive}
 The standard ``naive'' method of computing the length of an LCS is a
``bottom-up'' dynamic programming approach based on the following
recurrence for the length $L[i,j]$ of an LCS of $a_1a_2\ldots a_i$ and
$b_1b_2\ldots b_j$:
 \begin{equation}
 \label{eqn:naiverecur}
 L[i,j] =
    \left\{
    \begin{tabular}{ll}
    0 & if $i=0$ or $j=0$ \\
    $L[i-1,j-1]+1$ & if $i,j>0$ and $a_i=b_j$ \\
    $\max\{L[i-1,j],L[i,j-1]\}$ & otherwise
    \end{tabular}
    \right.
 \end{equation}
 We refer to the $L[i,j]$ value as the rank of $[i,j]$, and we call
$[i,j]$ a match if $a_i=b_j$.
 In $O(mn)$ time, one may fill an array with all the values of
$L[i,j]$ for $0\leq i\leq m \wedge 0\leq j\leq n$, and the length $l$
of an LCS is read off from $L[m,n]$.  The same time bound also
suffices to produce a single LCS by a ``backtracing'' approach
starting from position $[m,n]$ of the array.  At each stage we just
step from position $[i,j]$ to a position $[i-1,j-1]$, $[i-1,j]$, or
$[i,j-1]$ that is responsible for the setting of $L[i,j]$ as
per~(\ref{eqn:naiverecur}); each match encountered generates a
character of the LCS (in reverse order).

The naive approach to generate all LCS embeddings~\cite{AhoU1995}
would be to extend the backtracing method as follows.  (To generate
all distinct LCSs, one could generate all embeddings and remove
duplicate LCSs.)  At each step, we would consider three possibilities
(and continue recursively); from position $[i,j]$, we could add a
character to the LCS and move to $[i-1,j-1]$ if $[i,j]$ is a match,
and we could move to $[i-1,j]$ or $[i,j-1]$ if the rank there equals
$L[i,j]$ (without adding a character to the LCS and regardless of
whether $[i,j]$ is a match).  Whenever we reach $[0,0]$, we can print
out an LCS embedding.  We could make some simple improvements such as
stopping each backtrace path at any position of rank 0, but this will
not change the basic degree of inefficiency as expressed in the
theorem below.  (Note that output size is always at least 1 rather
than 0, because the empty string $\varepsilon$ is a common subsequence
of any pair of input sequences.)

\begin{theorem}
 \label{thm:naive}
 The naive method of generating all LCS embeddings (or all LCSs) may
require time exceeding the output size by a factor of
$\Theta(\binomial{n+m}{m})$ in the worst case.
 \end{theorem}

\begin{proof}
 For the upper bound, consider the ``normalized'' time $N[i,j]$,
representing the time to complete the naive backtrace procedure from
position $[i,j]$, divided by $\max\{1,L[i,j]\}$.  An induction
argument shows that there are positive constants $c$ and $d$ such that
$N[m,n]\leq c\binomial{n+m}{m}-d$.  It is easy to choose constants and
obtain $N[i,j]\leq c\binomial{i+j}{i}-d$ for any $[i,j]$ with $i\leq1$
or $j\leq1$.  Included in this result is that $N[i,j]\leq
c\binomial{i+j}{i}-d$ for $i+j\leq3$.  We then complete the induction
by showing, for an arbitrary $m,n\geq2$, that
 $N[m,n]\leq c\binomial{n+m}{m}-d$ given that
 $N[i,j]\leq c\binomial{i+j}{i}-d$ for $i+j<n+m$.  For this final
step, we perform the following case analysis, with $l$ denoting the
rank of $[m,n]$.

Case I: $[m,n]$ is not a match.  Then
 $N[m,n] \leq N[m-1,n]+N[m,n-1]+O(1)$.
 (It is easy to see that this relationship holds if $N$ represents
ordinary time, since the traceback from $[m,n]$ does not need to add
anything to the outputs generated in the tracebacks from $[m-1,n]$ and
$[m,n-1]$.  The relationship then holds for normalized time, since the
ranks of $[m-1,n]$ and $[m,n-1]$ can be no higher than $l$.)  The
induction step can be completed by invoking the induction hypothesis
and using the familiar identity
 \begin{equation}
 \label{eqn:pascaltri}
 \binomial{r-1}{k-1} + \binomial{r-1}{k} = \binomial{r}{k} \ .
 \end{equation}

Case II: $[m,n]$ is a match.  The following three subcases cover all
possibilities (albeit with overlap between cases IIA and IIB).

Case IIA: $[m-1,n]$ is not of rank $l$.  Then, we have
 $N[m,n]\leq N[m-1,n-1]+N[m,n-1]+O(1)$.
 (Here, this relationship would not be valid with $N$ representing
ordinary time, because every output produced in the traceback from
$[m-1,n-1]$ must be augmented with an additional character
corresponding to the match at position $[m,n]$.  But with normalized time,
the relationship can be justified as follows.  $[m-1,n-1]$ is
of rank $l-1$, so $(l-1)N[m-1,n-1]$ is an upper bound on the amount of
time spent in the backtrace from $[m-1,n-1]$.  Furthermore,
$N[m-1,n-1]$ is an upper bound on the number of outputs produced in
the backtrace from $[m-1,n-1]$ and therefore on the amount of extra
time appending a single extra character to each output.  Since
$[m,n-1]$ is also of rank at most $l$ and we need not add anything to
the outputs of the backtrace from $[m,n-1]$, the desired relationship
holds.)  We can now complete the induction step in a similar fashion
to Case I.

Case IIB: $[m,n-1]$ is not of rank $l$.  This case is completely
analogous to Case IIA.

Case IIC: $[m-1,n]$ and $[m,n-1]$ are both of rank $l$.
 Then
 \begin{equation}
 \label{eqn:match}
 N[m,n] \leq N[m-1,n]+N[m-1,n-1]+N[m,n-1]+O(1) \ .
 \end{equation}
 Furthermore, since $[m,n]$ is a match, $[m-1,n-1]$ is at rank $l-1$,
which is lower than the ranks of $[m-1,n]$ and $[m,n-1]$.  Thus,
 \begin{equation}
 \label{eqn:match-up}
 N[m-1,n] \leq N[m-2,n]+N[m-2,n-1]+O(1) \\
 \end{equation}
 \begin{equation}
 \label{eqn:match-left}
 N[m,n-1] \leq N[m,n-2]+N[m-1,n-2]+O(1)
 \end{equation}
 Combining, Equations~\ref{eqn:match}, \ref{eqn:match-up},
and~\ref{eqn:match-left}, we have
 $$
 N[m,n] \leq N[m-2,n]+N[m-2,n-1]+N[m-1,n-1]+N[m-1,n-2]+N[m,n-2]+O(1)
 $$
 Now we are again able to complete the induction step as in Case I,
using Equation~\ref{eqn:pascaltri} several times.

From the case analysis, we have concluded that the normalized time
$N[m,n]$ is $O(\binomial{n+m}{m})$.  Since the true output size of
listing even just distinct LCSs is at least $l=L[m,n]$, the overhead
of the naive algorithm is $O(\binomial{n+m}{m})$.

For the lower bound, note that an overhead of
$\Theta(\binomial{n+m}{m})$ is achievable by simply choosing sequences
with no matches.  Furthermore, even if we make the backtracing
procedure less naive by printing outputs whenever we hit a node of
rank 0, we still would spend
 $\Omega(\binomial{n+m-2}{m-1})=\Omega(\binomial{n+m}{m})$
 time for a pair of input strings in which the only match is at
$[1,1]$, while the true output size would be 1 even to list all
embeddings of all LCSs.
 \qed
 \end{proof}

We can now give a simple expression for the worst-case overhead of the
naive algorithm on two input strings of equal length.  This result
follows from expressing $\binomial{2n}{n}$ from
Theorem~\ref{thm:naive} as $\frac{(2n)!}{(n!)^2}$ and using
Equation~\ref{eqn:Stirling}.

\begin{corollary}
 \label{cor:naive-m=n}
 For two input strings of length $n$,
 naively
 generating all LCS embeddings (or all LCSs) may require time
exceeding the output size by a factor of $\Theta(4^n/\sqrt{n})$ in the
worst case.
 \qed
 \end{corollary}

Finally, we can recast the result for the case in which the total
number of characters in the two input strings is fixed, but the
lengths of the individual strings are not.

\begin{corollary}
 \label{cor:naive-t}
 For two input strings of total length $t$,
 naively
 generating all LCS embeddings (or all LCSs) may require time exceeding
the output size by a factor of $\Theta(2^t/\sqrt{t})$ in the worst
case.
 \end{corollary}

\begin{proof}
 From Theorem~\ref{thm:naive}, the worst-case overhead is based on the
maximum value of $\binomial{t}{m}$, which is
$\binomial{t}{\ceil{t/2}}$.  Then we proceed as for
Corollary~\ref{cor:naive-m=n}.
 \qed
 \end{proof}

\section{Conclusion}
 \label{sec:conc}
 We have seen that the maximum number of distinct longest common
subsequences for fixed input length is much less than the maximum
number of LCS embeddings, which is much less than the maximum number
of embeddings (including duplicates) obtained by generating embeddings
by the standard method.  Thus, it is much more efficient to generate
all distinct LCSs or all LCS embeddings in time proportional to the
output size than to use the standard method of generating LCS
embeddings.

\bibliographystyle{hplain}
\bibliography{sources}

\end{document}